\documentclass[aps, showpacs, amsmath,twocolumn,amssymb,floatfix,nofootinbib,
superscriptaddress,prl]{revtex4}
\usepackage{graphicx} 

\usepackage{color}
\definecolor{Blue}{rgb}{0.3,0.3,0.9}
\usepackage[T1]{fontenc}
\usepackage{ae,aecompl}
\usepackage{pstricks}
\usepackage{pstricks-add}
\usepackage{epsfig}
\usepackage{epstopdf}

\newcommand{\vect}[1]{\boldsymbol{#1}}

\begin{document}


\title{Electrical control of the Kondo effect in a helical edge liquid}


\author{Erik Eriksson}
\affiliation{Department of Physics, University of Gothenburg, SE-412 96 Gothenburg, Sweden}
\author{Anders Str\"om}
\affiliation{Department of Physics, University of Gothenburg, SE-412 96 Gothenburg, Sweden}
\author{Girish Sharma}
\affiliation{Centre de Physique Th\'{e}orique, Ecole Polytechnique, FR-91128 Palaiseau Cedex, France}
\author{Henrik Johannesson}
\affiliation{Department of Physics, University of Gothenburg, SE-412 96 Gothenburg, Sweden}


\begin{abstract}
Magnetic impurities affect the transport properties of the helical edge states of quantum spin Hall insulators by causing single-electron backscattering. 
We study such a system in the presence of a Rashba spin-orbit interaction induced by an external electric field, showing that this can be used to control the Kondo temperature, as well as the correction to the conductance due to the impurity. Surprisingly, for a strongly anisotropic electron-impurity spin exchange, Kondo screening may get obstructed by the presence of a non-collinear spin interaction mediated by the Rashba coupling. This challenges the expectation that the Kondo effect is stable against time-reversal invariant perturbations. 

\end{abstract}

\pacs{71.10.Pm, 72.10.Fk, 85.75.-d}


\maketitle


{\em Introduction.}
The discovery that HgTe quantum wells support a quantum spin Hall (QSH) state  \cite{Konig} has set off an avalanche of studies
addressing the properties of this novel phase of matter \cite{Review}. A key issue has been to determine the conditions for stability of the 
current-carrying states at the edge of the sample as this is the feature that most directly
impacts prospects for future applications in electronics/spintronics. In the simplest picture of a QSH system the 
edge states 
are {\em helical}, with counter-propagating electrons carrying opposite spins. By time-reversal invariance electron transport then becomes ballistic, 
provided that the electron-electron (e-e) interaction is sufficiently well-screened so that higher-order scattering 
processes do not come into play \cite{Wu,Xu}. 

The picture gets an added twist when including effects from magnetic impurities, contributed by 
dopant ions or electrons trapped by potential inhomogeneities. Since an edge
electron can backscatter from an impurity via spin exchange, time-reversal invariance no longer protects the 
helical states from mixing. In addition, correlated two-electron \cite{Meidan} and inelastic single-electron processes \cite{Schmidt, Lezmy} 
must now also be accounted for. 
As a result, at high temperatures $T$ electron scattering off the impurity 
leads to a $\ln (T)$ correction of the conductance at low frequencies $\omega$ \cite{M}, which, however,
vanishes in the dc limit $\omega \rightarrow 0$~\cite{T}. At low $T$, for weak
e-e interactions, the quantized edge conductance $G_0 = e^2/h$ is restored as $T\rightarrow 0$ with power laws
distinctive of a helical edge liquid. For strong interactions the edge liquid freezes into an insulator at $T=0$, with thermally induced transport
via tunneling of fractionalized charge excitations through the impurity \cite{M}.  

A more complete description of edge transport in a QSH system must include also the presence of a Rashba spin-orbit
interaction. This interaction, which can be tuned by an external gate voltage, is a built-in feature of a quantum well \cite{WinklerBook}. 
In fact, HgTe quantum wells exhibit some of the
largest known Rashba couplings of any semiconductor heterostructures \cite{Buhmann}. As a consequence, spin is no longer
conserved, contrary to what is assumed in the minimal model of a QSH system \cite{BHZ}. However,
 since the Rashba interaction preserves time-reversal invariance,  Kramers' theorem guarantees that the edge states are still connected via a time-reversal transformation  
 ("Kramers pair") \cite{Kane}. Provided that
 the Rashba interaction is spatially uniform and the e-e interaction is not too strong, this ensures the robustness of the helical
 edge liquid \cite{Strom}.

What is the physics with both Kondo {\em and} Rashba interactions present? In this paper we
address this question with a renormalization group (RG) analysis as well as a linear-response and rate-equation approach.
Specifically, we predict that the Kondo temperature $T_K$ $-$ which sets the
scale below which the electrons screen the impurity $-$ can be controlled by varying the strength of the Rashba interaction. 
Surprisingly, for a strongly anisotropic Kondo exchange, a non-collinear spin interaction mediated  by the Rashba coupling 
becomes relevant (in the sense of RG) and competes with the Kondo screening. This challenges the expectation that the Kondo 
effect is stable against time-reversal invariant perturbations \cite{Meir}. 
Moreover, we show that the impurity contribution to the dc conductance at temperatures $T\!>\!T_K$ can be switched on and 
off by adjusting the Rashba coupling. With the Rashba coupling being tunable by a gate voltage,  
this suggests a new inroad to control charge transport at the edge of a QSH device. 

{\em Model.}
To model the edge electrons, we introduce the two-spinors $\Psi^T = \big( \psi_{\uparrow}, \psi_{\downarrow} \big)$, where $ \psi_{\uparrow} \, (\psi_{\downarrow})$
annihilates a right-moving (left-moving) electron with spin-up (spin-down) along the growth direction of the quantum well. Neglecting e-e interactions,
the edge Hamiltonian can then be written as
\begin{eqnarray} \label{H}
H &=& v_F \int \textrm{d}x \ \Psi^{\dagger}(x) \left[ -\textrm{i} \sigma^{z} \partial_x \right] \Psi (x) + \nonumber \\
&& + \alpha \int \textrm{d}x \ \Psi^{\dagger}(x) \left[ -\textrm{i} \sigma^{y} \partial_x \right] \Psi (x) + \\
&& + \Psi^{\dagger} (0) \left[  J_x \sigma^x S^x + J_y \sigma^y S^y + J_z \sigma^z S^z \right] \Psi(0), \nonumber
\end{eqnarray}
with $v_F$ the Fermi velocity parameterizing the linear kinetic energy. The second term encodes the Rashba interaction of strength 
$\alpha$, 
with the third term being an antiferromagnetic Kondo interaction 
between electrons (with Pauli matrices $\sigma^{i}, i = x,y,z$) and a spin-1/2 magnetic impurity (with Pauli matrices $S^{i}, i=x,y,z$)
at $x=0$. The spin-orbit induced magnetic anisotropy for an impurity at a quantum well interface \cite{Ujsaghy} implies 
that $J_x = J_y \neq J_z$ \cite{Zitko}. Unless otherwise stated, we use $\hbar \,  = k_B \equiv 1$.

The Rashba term in Eq. (\ref{H}) can be absorbed into the kinetic term by a a spinor rotation
$\Psi \rightarrow \Psi' = \mathrm{e}^{-\mathrm{i}\sigma^x \theta/2} \Psi$ \cite{VO}. By rotating also the impurity spin, $\vect{S} \rightarrow \vect{S}^{\prime} = \mathrm{e}^{-\mathrm{i}S^x \theta/2}\vect{S}\mathrm{e}^{\mathrm{i}S^x \theta/2}$, one obtains $H= H_0^{\prime}+ H_K^{\prime}$, with
\begin{eqnarray} 
H_0^{\prime} &=& v_{\alpha} \int \textrm{d}x \ \Psi'^{\dagger}(x) \left[ -\textrm{i} \sigma^{z'} \partial_x \right] \Psi' (x) \label{H0}\\
H_K^{\prime} &=& \Psi'^{\dagger} (0) [J_x \sigma^{x} S^{x} + J'_y \sigma^{y'} S^{y'} + J'_z \sigma^{z'} S^{z'} \nonumber \\
&& \qquad  \ \ \ + J_E ( \sigma^{y'} S^{z'} +  \sigma^{z'} S^{y'}) ] \Psi'(0) , \label{HK}
\end{eqnarray}
where $J'_y = J_y \cos^2 \theta + J_z \sin^2 \theta$, $J'_z = J_z \cos^2 \theta + J_y \sin^2 \theta$ and 
$J_E = (J_y - J_z )\cos \theta \sin \theta$. The Rashba rotation angle $\theta$ is determined through
$\cos \theta = v_F / v_{\alpha}$, $\sin \theta = \alpha / v_{\alpha}$ and $ v_{\alpha} = \sqrt{v_F^2 + \alpha^2}$. Note that the
spin in the rotated basis is quantized along the $z'$-direction which forms an angle $\theta$ with the $z$-axis. 
Also note that the Kondo interaction in the new basis not only becomes spin-nonconserving, but also picks up a non-collinear term
for $J_y \neq J_z$. 

Including e-e interactions, and assuming a band filling incommensurate with the lattice \cite{Review}, 
time-reversal invariance constrains the possible scattering channels in the rotated basis to dispersive $(\sim\!g_d)$ and forward $(\sim~g_f)$ scattering, in addition to correlated two-particle backscattering $(\sim\!g_{bs})$ \cite{Wu, Xu} and inelastic single-particle backscattering
$(\sim\!g_{ie})$ \cite{Schmidt,Lezmy,Budich} at the impurity site. Adding the corresponding interaction terms to $H_0^{\prime}$ and $H_K^{\prime}$
in (\ref{H0}) and (\ref{HK}), and employing standard bosonization
\cite{G}, the full Hamiltonian for the edge liquid can now be expressed as a free boson model, $(v/2) \!\int \textrm{d}x \!\left( (\partial_x\varphi)^2 +
(\partial_x \vartheta)^2\right)$, with three local terms added at $x=0$:
\begin{eqnarray}
H_K^{\prime}\! \!&\!=\!& \!\!\frac{A}{\kappa}\!\cos(\sqrt{4\pi K}\varphi)\! +\! \frac{B}{\kappa} \!\sin(\sqrt{4\pi K}\varphi)\! + \!
\frac{C}{\sqrt{K}}\partial_x\vartheta \label{KondoBoson} \\   
H_{bs}^{\prime} \!&\!=\!&\! \frac{g_{bs}}{2(\pi \kappa)^2} \cos(\sqrt{16\pi K}\varphi) \label{BackBoson} \\
H_{ie}^{\prime} \!&\!=\!&\! \frac{g_{ie}}{2\pi^2\sqrt{K} }:(\partial_x^2 \vartheta) \cos(\sqrt{4\pi K}\varphi) :      \label{IEBoson}
\end{eqnarray}
Here $\varphi$ is a nonchiral Bose field with $\vartheta$ its dual, $v\partial_x \vartheta = \partial_t \varphi$ with $v \!=\![(v_{\alpha} \!+ \!g_f/\pi)^2\!-\!(g_d/\pi)^2]^{1/2}$, $K =  [(\pi v_{\alpha} \!+ \!g_f - g_d)/(\pi v_{\alpha} \!+ \!g_f + g_d)]^{1/2}$
and $\kappa \, \approx v_F/D$ is the edge state penetration depth acting as short-distance cutoff with $D$ the band width,
and $:...:$ denotes normal ordering. In $H'_K$ we have defined
$A=J_x S^{x} /\pi$, $B=(J'_yS^{y'}+J_ES^{z'})/\pi$ and $C=(J'_zS^{z'}+J_E S^{y'})/\pi$.

{\em Kondo temperature.} 
The bosonized theory is tailor-made for a perturbative RG analysis, allowing us to determine the temperature scale below which
the edge electrons couple strongly to the impurity. We first note that the backscattering term in (\ref{BackBoson}) is that of the well-known boundary-sine Gordon model.
For $K\!<\!2/3$ it dominates over the inelastic backscattering in (\ref{IEBoson}), and turns relevant for $K\!<\!1/4$ with a weak to strong-coupling crossover
at a temperature $T_{bs} \approx Dg_{bs}^{1/(1-4K)}$ \cite{KaneFisher}.
As a consequence, the enhancement of backscattering 
as $T \rightarrow 0$ results in a zero-temperature insulating state when $K<1/4$. 

Turning to the Rashba-rotated Kondo interaction $H_K^{\prime}$ in Eq. (\ref{KondoBoson}), we obtain for its one-loop RG equations:  
\begin{gather}
\partial_l \tilde{J}_x = (1\!-\!K)\tilde{J}_x + \nu K(\tilde{J}_y^{\prime}\tilde{J}_z^{\prime}\!-\!\tilde{J}_{E1} \tilde{J}_{E2})\nonumber \\
\partial_l \tilde{J}_y^{\prime} =  (1\!-\!K)\tilde{J}_y^{\prime} + \nu K \tilde{J}_x \tilde{J}_z^{\prime}, \ \ \partial_l \tilde{J}_z^{\prime} = \nu K\tilde{J}_x \tilde{J}_y^{\prime},  \label{RGequations}\\
\partial_l \tilde{J}_{E1}=  (1\!-\!K)\tilde{J}_{E1} -\nu K\tilde{J}_x \tilde{J}_{E2}, \  \  \partial_l \tilde{J}_{E2} \!=\! -\nu K\tilde{J}_x \tilde{J}_{E1}, \nonumber
\end{gather} 
\noindent with the ''tilde'' indicating that the couplings depend on the renormalization length $l$, and where $\nu \equiv 1 /(\pi v)$. The two terms proportional to $J_E$ in Eq. (\ref{KondoBoson}) flow individually under RG, with the corresponding renormalized coupling constants here denoted $\tilde{J}_{E1}$ and $\tilde{J}_{E2}$. In deriving Eqs. (\ref{RGequations}) we have used that higher-order contributions involving an intermediate process governed by $H_{bs}^{\prime}$ or $H_{ie}^{\prime}$ are suppressed, since these transfer spin or energy incompatible with $H_K^{\prime}$. In a recent work \cite{M2}, Kondo scattering without Rashba interaction was studied, and different physics in the regime $\nu J_z\ge 2K$ was found, not accessible perturbatively in $\nu J_z$.  Since its realization in an HgTe quantum well requires anomalously weak screening of the e-e interaction we do not consider this regime here.

The role of the Rashba rotation in Eqs.\ (\ref{RGequations}) is both to determine the bare values $\tilde{J}'_{y,z}(l=0)\equiv J'_{y,z}$ and to introduce the non-collinear couplings $\tilde{J}_{E1,E2}$. To explore the outcome, we first examine the case of a strongly screened e-e interaction, $K\approx 1$. For this case, the first-order terms of Eq.~(\ref{RGequations}) can be neglected and $\tilde{J}_{E1}=\tilde{J}_{E2}=\tilde{J}_E$, since their scaling equations will be identical. In this limit, $\tilde{J}_E$ quickly flows to zero. We take the Kondo temperature $T_K$ to be the value of $T=D\exp(-l)$ where one of the couplings in Eq. (\ref{RGequations}) first grows past $1/(\nu K)$, making the renormalized $H_K^{\prime}$ in Eq.~(\ref{KondoBoson}) dominate the free theory. For $K\approx1$ we then see that
\begin{equation}  \label{TKmod}
T_K \approx D \exp\left(-\frac{1}{\nu J_x}\frac{\operatorname{arcsinh}(\zeta)}{\zeta}\right),
\end{equation}

\noindent where $\zeta=\sqrt{(J_z/J_x)^2-1}$ is an anisotropy parameter~\cite{M}. Here the $\theta$ dependence lies predominantly in $\nu$. Note that Kondo temperatures modified 
by spin-orbit couplings, as in (\ref{TKmod}), or by spin-dependent hopping, have recently been proposed also for ordinary conduction electrons \cite{Schuricht,Feng,Zitko2,Zarea, Isaev}.

In the opposite limit of a strong e-e interaction, the second-order terms of the scaling equations can be neglected, as long as $1-K\gg \tilde{J} \nu K$, for all $\tilde{J} = \tilde{J}_x, \tilde{J}'_y, \tilde{J}_{E1}$. The scaling equations in this limit reduce to $\partial_l \tilde{J}=(1-K) \tilde{J}$, with solutions $\tilde{J}=J e^{(1-K)l}$. With $l=\ln(D/T)$, one can now use the $\tilde{J}=1/(\nu K)$ criterion to find the Kondo temperature
\begin{equation}
T_K \approx D \left(J_{max}\,\nu K\right)^{1/(1-K)},
\end{equation}
\noindent where $J_{max}=\max[J_x,J'_y,J_{E}]$. 

\begin{figure}[t!]
	\begin{center}
		\includegraphics[width=1.0\columnwidth]{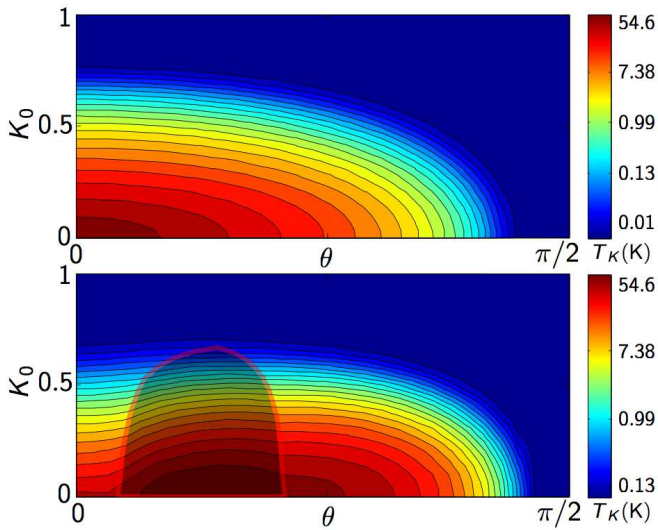}
		\caption{The Kondo temperature $T_K$ as a function of the Rashba angle $\theta$ and the ordinary Luttinger parameter $K_0$. The $T_K$ scale is logarithmic and red and blue color indicates high and low $T_K$, respectively. Top:  $J_x = J_y \geq J_z$ (here, $J_x/a_0=J_z/a_0=10$meV), bottom: $J_x  = J_y <J_z$ (here, $J_x/a_0=5$ meV and $J_z/a_0=50$ meV). In the shaded area, $\tilde{J}_{E1}$ dominates the perturbative RG flow, hence obstructing singlet formation.}
		\label{KondoPlot}
	\end{center} 
\end{figure}
In Fig.\ (\ref{KondoPlot}) we exhibit $T_K$ for both ''easy-plane'' and ''easy-axis'' Kondo interaction. To isolate the effect of the Rashba interaction from that of the e-e interaction we choose to plot $T_K$ as a function of $\theta$ and $K_0$, with $K_0 \equiv K(\theta$$=$$0)$ the ordinary 
Luttinger parameter. For $|J_E| > \left|J_x\right|,\left|J'_y\right|$, the non-collinear term $\sim \sigma^{ y'}S^{z'}$ in Eq. 
(\ref{HK}) dominates the RG flow for values of $K$ in the shaded ''dome'' (the size of which is set by the ratio $J_z/J_{x,y}$). 
%
As this term disfavors a spin singlet,  Kondo screening will be 
obstructed in the corresponding interval of Rashba couplings \cite{domefootnote}. 
This runs contrary to the expectation that a spin-orbit interaction
does not impair the Kondo effect \cite{Meir,Malecki}. However, this expectation is rooted in a noninteracting quasiparticle picture which breaks down in one dimension. Instead a  
Luttinger liquid is formed, with strongly correlated electron scattering \cite{G}. As suggested by our RG analysis, when this scattering gets enhanced with lower values of $K$, 
it boosts the effect of the non-collinear spin interaction that works against the Kondo screening.

{\em Conductance at low temperatures.} Away from the ''dome'' in Fig.\ (\ref{KondoPlot}), the Rashba-rotated Kondo interaction easily sustains a Kondo temperature $T_K$ below which the impurity gets screened. When $K>1/4$ and two-particle backscattering is RG-irrelevant, there is no correction $\delta G$ to the conductance at zero temperature: As explained by Maciejko {\em et al.} \cite{M}, the topological nature of the QSH state implies a "healing" of the edge after the impurity has been effectively removed by the Kondo screening. For finite $T\!\ll \!T_K$, the leading correction $\delta G$ is generated by either $H_{bs}^{\prime}$ or $H_{ie}^{\prime}$, whatever operator has the lowest scaling dimension: For $1/4 < \!K <\! 2/3 \, (K > 2/3)$ $H_{bs}^{\prime}\, (H_{ie}^{\prime})$ dominates, with $\delta G_{bs} \sim (T/T_K)^{8K-2}$  \cite{M} $(\delta G_{ie} \sim (T/T_K)^{2K+2}$ \cite{Schmidt,Lezmy}). 
The picture changes dramatically for $K<1/4$. Now $H_{bs}$ turns RG-relevant, with $g_{bs}$ entering a strong-coupling regime below the crossover-temperature $T_{bs}$, implying zero conductance at $T=0$. At finite $T \ll T_{bs}$, instanton processes restore its finite value, with $G \sim (T/T_{bs})^{2(1/4K-1)}$~\cite{M}. To leading order this regime is blind to the Rashba interaction. 

{\em Conductance and currents at high temperatures.}
When $T > \mbox{max}(T_K, T_{bs})$, scattering from $H_K^{\prime}$ as well as from $H_{bs}^{\prime}$ and $H_{ie}^{\prime}$ remain weak, and transport properties can be obtained perturbatively. We here focus on the correction $\delta G$ to the conductance due to $H_K^{\prime}$, noting that the contributions from $H_{bs}^{\prime}$ and $H_{ie}^{\prime}$ decouple and are insensitive to the strength  of the Rashba interaction.

The current operator $\hat{I}$ takes the form $\hat{I}= (e/2)\,\partial_t \,( \Psi'^{\dagger} \sigma^{z'} \Psi')$ in the rotated basis, since the components of the rotated spinor define new right and left movers. After the unitary transformation $U= \mathrm{e}^{\mathrm{i}\sqrt{\pi}\lambda\varphi(0)S^{z \prime}}$, which removes the $J'_z$-term when $\lambda =  J'_z / \pi v \sqrt{K}$, the bosonized part $\delta \hat{I}$ of the current operator due to $H_K^{\prime}$ is 
\begin{align} \label{dI2} \hspace{-0.5cm} 
\delta \hat{I} & =  \frac{\mathrm{i}e}{2\pi\kappa}  \big[ \!\sum_{j=\pm}\! A_j  e^{\mathrm{i\sqrt{\pi}}(2\sqrt{K}-j\lambda)\varphi} S^j + \mathrm{i} A_0 \, e^{\mathrm{i}\sqrt{4\pi K}\varphi} S^z \big]  \nonumber \\
 & \hspace{6cm}  + \mbox{H.c.}  
\end{align} 
where $A_{\pm} = (1/2)(J_x \pm J'_y)$, and $ A_0 = J_E/2$. Using the Kubo formula to calculate the conductance correction  
$\delta G (\omega)$ at a frequency $\omega$ in the limit $J^2 \ll \omega \ll T $, with $J^2 = J_x^2, J_y'^2, J_E^2$,
we then find to $\mathcal{O}(J^2)$
\begin{equation}\label{dG} 
\delta G  = - \, \frac{e^2}{\hbar}\sum_{j= -1}^{+1} A_j^2  \cdot  F(2\sqrt{K} -j\lambda) \cdot    (2\pi T/D)^{2(\sqrt{K}-j\lambda/2)^2-2} ,
\end{equation}
which, in this limit, is independent of $\omega$. Here $F(x) = [ \Gamma(x^2 /4)]^2 / [4\pi (\hbar v)^2 \Gamma(x^2/2)] $.
At zero Rashba coupling,  $\theta \!=\! 0$, Eq.~(\ref{dG}) reproduces the finding in Refs.~\onlinecite{M,T}. 
By replacing the bare couplings by renormalized ones, the result in Eq. (\ref{dG}) can be RG-improved to numerically obtain $\delta G$ to all orders in perturbation theory in a leading-log approximation.
At $\theta \! =\! 0$ this gives $\delta G \sim \ln (T)$, in agreement with Ref. \onlinecite{M}.

As stressed in Ref.~\onlinecite{T}, the use of the Kubo formula rests on a perturbation expansion (in our case assuming that $J^2 \ll \omega$) which breaks down as
$\omega \to 0$. To study the scaling of $\delta G$ in the dc limit we will instead fall back on a rate equation approach. The details of this calculation are provided in the Supplemental Material, and we here only give the main results.
In the dc limit, i.e.  $\omega \ll J^2 \ll T$, the conductance correction becomes
\begin{displaymath} \label{rategdc} \hspace{-1cm}
\ \ \  \ \  \ \ \ \delta G =  -  \frac{e^2 }{2T} 
\Big[ \frac{ 4\gamma_0 \gamma'_0\! +\!(\gamma_0\!+\!\gamma'_0)(\gamma_0^E \!+\!\tilde{\gamma}_0^E ) \!+\! 
\tilde{\gamma}^E_0\gamma_0^E}{\gamma_0\!+\!\gamma'_0\! +\! \tilde{\gamma}_0^E } \Big]  
\end{displaymath}
with $\gamma_0 \sim (J_x+J_y^{\prime})^2 T^{2(\sqrt{K}-\lambda/2)^2-1}, \gamma_0^{\prime} \sim (J_x-J_y^{\prime})^2 T^{2(\sqrt{K}+\lambda/2)^2-1}, \gamma_0^E \sim J_E^2 T^{2K-1}$, and $\tilde{\gamma}_0^E \sim J_E^2 T$, where, for brevity, we have omitted various $K$-dependent prefactors.  
Note that with $J_x = J_y$, the vanishing $\delta G$ becomes non-zero when turning on the Rashba interaction by an electric field. This suggests a means to manipulate the edge current by varying the bias of an external gate. 

To explore this possibility we have calculated the $\delta I$-$V$ characteristics, exploiting that in the rotated basis $H'_K$ can be treated as a tunneling Hamiltonian \cite{Mahan} and $\delta I$, corresponding to the tunneling current, is then obtained as in Ref.~\onlinecite{W}. When $ J^2 \ll \omega \ll  T, eV$ we find
\begin{eqnarray} \label{ivc}
\delta I\  \approx  - e \sum_{j=-1}^{+1}  \mbox{Im}\{ \, B(K_j + \mathrm{i} e V/ 2\pi T , K_j -\mathrm{i} e V / 2\pi T  )  \nonumber \\ 
 \times C_j (T/D)^{ 2K_j-1}   \sin[\pi(K_j -\mathrm{i} e V / 2\pi T )] / \cos (\pi K_j) \} \quad 
\end{eqnarray}
for $\delta I \equiv I - G_0 V$, with $K_j \equiv (\sqrt{K}-j\lambda/2)^2$, and $B$ the beta function. Here $C_{\pm1} \!= \!c_{\pm} (J_x \pm J'_y)^2$ and $C_{0} \!= \!c_0 J_E^2$, with $c_{\pm,0}$ constants depending on $K$, $\lambda$ and $\theta$. In Fig.~(\ref{Mnplot}) we plot this for parameter values given below.

\begin{figure}[hbt!]
	\begin{center}
		\includegraphics[width=1.0\columnwidth]{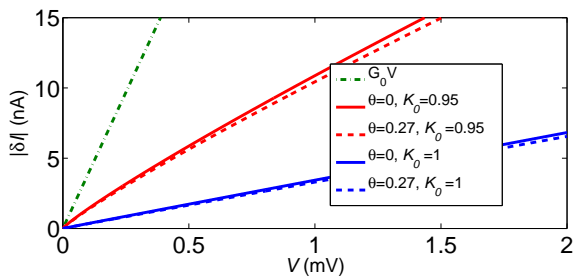} 
		\caption{The RG-improved current correction (\ref{ivc}) at $T=30$ mK as a function of applied voltage, for different values of $K_0$ and $\theta$. The dashed lines represent $\theta\approx 0.27$, corresponding to $\hbar\alpha=10^{-10}$ eVm. Other parameters are defined in the text. The QSH edge current $G_0V$ is plotted as a reference.} 
		\label{Mnplot}
	\end{center} 
\end{figure}

{\em Experimental realization.}
Given our result in Eq. (\ref{ivc}), is the Rashba-dependence of $\delta I$ large enough to be seen in an experiment? 
As a case study, let us consider an Mn$^{2+}$ ion implanted close to the edge of an HgTe quantum well \cite{Datta}.  
Mn$^{2+}$ has spin $S\!=\!5/2$, but, due to the large 
and positive single-ion anisotropy $\propto (S^z)^2$ at the quantum well
interface, the higher spin components freeze out in the sub-Kelvin range, leaving behind a spin-1/2 doublet \cite{Ujsaghy}.
Moreover, the single-ion anisotropy implies that the Kondo interaction with this effective spin-1/2 impurity is anisotropic, with $J_x = J_y = 3J_z = 3 J_I$,
where $ J_I $ is the isotropic bulk spin-exchange coupling \cite{Zitko}. Its value can be assessed from the sp-d exchange integrals for the bulk conduction electrons in Hg$_{1-x}$Mn$_x$Te \cite{Furdyna}. 
Close to the $\Gamma$-point of the Brillouin zone these integrals produce an antiferromagnetic exchange, $ J_I > 0$.  
With  
the Mn$^{2+}$ ion located within the penetration depth $\kappa$ from the edge,  a rough estimate yields an expected value of $ J_I/a_0 \approx 10$ meV, with $a_0$ the lattice constant. Turning to the Rashba coupling $\alpha$, gate controls have been demonstrated in the laboratory with $\hbar \alpha$ for an HgTe quantum well device running from $5\times 10^{-11}$ eVm to $1 \times 10^{-10}$ eVm as the bias of a top gate is varied from 2 V to -2 V \cite{Hinz}. As for the value of the interaction parameter $K_0$ in an HgTe quantum well, estimates range between 0.5 and 1 \cite{M,Strom,Hou,Strom2,TeoKane}, and depend on the geometry and composition of the heterostructure. Collecting the numbers, and putting $a_0 \approx 0.5$ nm \cite{Efros}, $v_F \approx 5.0 \times 10^5$ m/s \cite{Konig}, and $D \approx 300$ meV \cite{Konig}, we can use Eq.~(\ref{ivc}) to numerically plot the $\delta I$-$V$ characteristics for different values of $\alpha$ and $K_0$ \cite{footnote}, choosing $T=30$ mK ($>$$T_K$), see Fig.~(\ref{Mnplot}). As revealed by the graphs, the Rashba-dependence of $\delta I$ should allow for an experimental test \cite{footnote2}. 

{\em Concluding remarks.}
We have studied the combined effect of a Kondo and a Rashba interaction at the edge of a quantum spin Hall system. The interplay between an anisotropic Kondo exchange and the Rashba interaction is found to  result in a non-collinear electron-impurity spin interaction. A perturbative RG analysis indicates that this interaction may block the Kondo effect when the electron-electron interaction is weakly screened. We conjecture that this surprising result $-$ challenging a time-honored expectation that the Kondo effect is blind to time-reversal invariant perturbations \cite{Meir} $-$ is due to the breakdown of single-particle physics in one dimension. It remains a challenge to unravel the microscopic scenario behind this intriguing phenomenon. In the second part of our work we derived expressions showing how charge transport at the edge is influenced by the simultaneous presence of a magnetic impurity and a Rashba interaction. A case study suggests that the predicted current-voltage characteristics should indeed be accessible in an experiment. Most interestingly, its manifest dependence on the gate-controllable Rashba coupling breaks a new path for charge control in a helical electron system. 

{\em Acknowledgments.} It is a pleasure to thank S. Eggert, D. Grundler, C.-Y. Hou, G. I. Japaridze, P. Laurell, T. Ojanen, and D. Schuricht for valuable discussions. We also thank F. Cr\'epin and P. Recher for drawing our attention to an erroneous expression for the current operator in an earlier version of this paper \cite{erratum}. This research was supported by the Swedish Research Council (Grant No. 621-2011-3942)  and by STINT (Grant No. IG2011-2028).

\newpage

\widetext

\appendix


\section{Supplemental Material}

In this supplementary material we derive the conductance correction $\delta G (\omega)$ as a function of the frequency $\omega$, using a rate equation approach in the spirit of Tanaka \textit{et al.}~[9]. Since the rotated spin-up ($\uparrow'$) and spin-down ($\downarrow'$) states define right and left movers, the current operator takes the form $\hat{I}= (e/2)\,\partial_t \,(\Psi'^{\dagger} \sigma^{z'} \Psi' )$ in the rotated basis.  A voltage $V = V_0  e^{-\mathrm{i}\omega t} $ adds a term
$H_V = (eV/2) \int \mathrm{d}x \,\,\Psi'^{\dagger}  \sigma^{z'}  \Psi'  $
to the Hamiltonian. A rate equation can now be constructed for the impurity spin,
\begin{equation} \label{re}
\partial_t P^i_{\uparrow'} = (\gamma_+ + \gamma'_- + \tilde{\gamma}^E_0) P^i_{\downarrow'} - (\gamma_- + \gamma'_+ + \tilde{\gamma}^E_0) P^i_{\uparrow'}, \tag{S.1}
\end{equation}
with $P^i_{\uparrow',\downarrow'}$ the probability of the impurity spin being in the $\uparrow'$ or $\downarrow'$ state, where $P^i_{\uparrow'} + P^i_{\downarrow'} = 1$. The solution is
\begin{equation} \label{pupp}
P^i_{\uparrow'} = \frac{1}{2} + \frac{e}{2T}\, \frac{\gamma_0-\gamma'_0}{2(\gamma_0 + \gamma'_0 + \tilde{\gamma}^E_0) - \mathrm{i}\omega }\, V_0  e^{-\mathrm{i}\omega t}  . \tag{S.2}
\end{equation}
The $\gamma$-parameters encode the various voltage-dependent spin-flip rates implied by $H'_{K}$ in Eq.~(3),
\begin{eqnarray} \label{HK2}
 \sigma^{\mp} S^{\pm}   \to  \gamma_{\pm}, \quad
 \sigma^{\pm} S^{\pm}   \to  \gamma'_{\mp} ,\quad
  \sigma^{\mp} S^{z'}   \to  \gamma^E_{\pm} ,\quad
      \sigma^{z'} S^{\pm}  \to  \tilde{\gamma}^E_0 . \nonumber
\end{eqnarray}
Here $\gamma_{\pm} = \gamma_0 \Lambda_{\pm}, \gamma^{\prime}_{\pm} = \gamma^{\prime}_0 \Lambda_{\pm}, \gamma^E_{\pm} = \gamma^E_0 \Lambda_{\pm}$,
with $\Lambda_{\pm} \equiv 1\pm eV/2T$ for $eV\ll T$, where the rates $\gamma_0$, $\gamma'_0$, $\gamma_0^E$ and $\tilde{\gamma}_0^E$ are determined below.

The current correction $\delta I = I - G_0 V$ due to the impurity is given by
\begin{equation}
 \delta I =  -e \,\big( \gamma_+ P^i_{\downarrow'} -\gamma'_- P^i_{\downarrow'} + \gamma'_+  P^i_{\uparrow'}  
- \gamma_- P^i_{\uparrow'} + \gamma^E_+ /2 - \gamma^E_- /2  \big)  . \tag{S.3}
\end{equation} 
Combining Eqs.~(S.2) and (S.3) gives for the conductance correction, $\delta G(\omega) = \delta I / (V_0 e^{-\mathrm{i}\omega t})$,
\begin{equation} \tag{S.4}
\delta G(\omega)= -  \frac{e^2}{2T}   \Big[ \frac{(\gamma_0+\gamma'_0 + \gamma_0^E)\omega + \mathrm{i}8\gamma_0 \gamma'_0 + \mathrm{i}2(\gamma_0+\gamma'_0)(\gamma_0^E +\tilde{\gamma}_0^E ) + \mathrm{i}2\tilde{\gamma}_0\gamma_0^E}{\mathrm{i}2(\gamma_0+\gamma'_0 + \tilde{\gamma}_0^E) + \omega} \Big].
\end{equation}
In the dc limit, i.e.  $\omega \ll J^2 \ll T$, we then obtain
\begin{equation} 
\delta G (\omega \to 0)=  - \frac{e^2}{2T} 
\Big[ \frac{ 4\gamma_0 \gamma'_0\! +\!(\gamma_0\!+\!\gamma'_0)(\gamma_0^E \!+\!\tilde{\gamma}_0^E ) \!+\! 
\tilde{\gamma}^E_0\gamma_0^E}{\gamma_0\!+\!\gamma'_0\! +\! \tilde{\gamma}_0^E } \Big] .  \tag{S.5}
\end{equation}
The rates $\gamma_0$, $\gamma'_0$ and $\gamma_0^E$ are now determined by considering the regime $\gamma_0$,$\gamma'_0$,$\gamma_0^E$,$\tilde{\gamma}_0^E \ll \omega \ll T$, where Eq.~(S.4) gives
\begin{equation} \tag{S.6}
\delta G(\omega \gg \gamma)  = -\frac{e^2}{2T} \big(\gamma_0+\gamma'_0 + \gamma_0^E   \big)  .
\end{equation}
Comparing Eq.~(S.6) with the linear-response result in Eq.~(11) immediately gives $\gamma_0 \sim (J_x+J_y^{\prime})^2 T^{2(\sqrt{K}-\lambda/2)^2-1}$, $\gamma_0^{\prime} \sim (J_x-J_y^{\prime})^2 T^{2(\sqrt{K}+\lambda/2)^2-1}$, and $\gamma_0^E \sim J_E^2 T^{2K-1}$. 

Obtaining $\tilde{\gamma}_0^E $ requires some additional work, since the terms  $\sigma^{z'} S^- $ and $ \sigma^{z'} S^+ $ in $H'_K$, Eq.~(3), do not backscatter electrons. Hence the rate $\tilde{\gamma}_0^E$ does not enter the linear-response conductance result in Eq.~(11). To make progress one may introduce an auxiliary field coupling to the impurity instead of the electrons. A suitable choice is to apply a magnetic field $h=h_0  e^{-\mathrm{i}\omega t}$ to the impurity spin and obtain the spin-flip rates when $h \to 0$ using linear response. The equilibrium probabilities for the impurity spin are then
\begin{equation} \tag{S.8}
P^i_{\uparrow',\downarrow'} = \frac{e^{\pm\mu h / 2T}}{e^{\mu h / 2T} + e^{-\mu h / 2T} },
\end{equation}
and the spin-flip rates induced by $H'_K$ now correspond to
\begin{eqnarray} \label{HK2}
 \sigma^{\mp} S^{\pm}   \to  \gamma_{\pm}, \quad
 \sigma^{\pm} S^{\pm}   \to  \gamma'_{\pm} ,\quad
  \sigma^{\mp} S^{z'}   \to  \gamma^E_{0} ,\quad
      \sigma^{z'} S^{\pm}  \to  \tilde{\gamma}^E_{\pm} , \nonumber
\end{eqnarray}
with $\gamma_{\pm} = \gamma_0 \Lambda_{\pm}, \gamma^{\prime}_{\pm} = \gamma^{\prime}_0 \Lambda_{\pm}, \tilde{\gamma}^E_{\pm} = \tilde{\gamma}^E_0 \Lambda_{\pm}$,
where we now have $\Lambda_{\pm} \equiv 1\pm \mu h/2T$ in the limit $\mu h\ll T$. Since the ballistic conduction electrons are in equilibrium with the leads, the rate equation for the impurity spin can be written as
\begin{equation} \tag{S.9}
\partial_t P^i_{\uparrow'} =  \frac{1}{2} \big[-\gamma_- + \gamma_+    +\gamma'_+   - \gamma'_-   - \tilde{\gamma}^E_- + \tilde{\gamma}^E_+ \ \big] = \frac{\mu h}{2T} \big(\gamma_0+\gamma'_0 + \tilde{\gamma}_0^E \big) .
\end{equation}
This gives the "spin-flip susceptibility", $\chi \equiv \partial (\partial_t P^i_{\uparrow'}) /  \partial (\mu h)$,
\begin{equation} \tag{S.10}
\chi = \frac{1}{2T} \big(\gamma_0+\gamma'_0 + \tilde{\gamma}_0^E \big).
\end{equation}
The rate $\tilde{\gamma}_0^E$ can now be extracted from a linear response calculation. The operator $\partial_t S^{z'} $, given by $\partial_t S^{z'} =(\mathrm{i}\hbar)^{-1}\,\big[ S^{z'} , H'_K \big]$, becomes
\begin{equation} \label{dSz} \hspace{-0.5cm}
\partial_t S^{z'} \ =\  \frac{\mathrm{i}}{\hbar }\frac{1}{2\pi \kappa}\sum_{j=\pm} \left(\frac{J_x +j J'_y}{2} \right)   e^{\mathrm{i}(2\sqrt{K}-j\lambda)\varphi(0)} S^ j   \ -\ \frac{1}{\hbar}\frac{J_E}{2} \frac{1}{\pi \sqrt{K}}  : \partial_x \vartheta(0) e^{-\mathrm{i}\lambda \varphi(0)} : S^+   \ +\  h.c.  . \tag{S.11}
 \end{equation}
Calculating the "spin-flip susceptibility" $\chi$ using the Kubo formula, $\chi(\omega) = (\hbar \omega)^{-1} \int_{0}^{\infty} \mathrm{d} t \, e^{\mathrm{i}\omega t} \langle \big[ (\partial_t S^{z'})^{\dagger} (t) , \partial_t S^{z'}(0) \big] \rangle$, in the regime $\gamma_0$,$\gamma'_0$,$\tilde{\gamma}_0^E$,$\tilde{\gamma}_0^E \ll \omega \ll T$, we get
\begin{equation} 
\chi \  =\   \, \frac{1}{\hbar} \sum_{j= \pm} \Bigg[  \frac{J_x + jJ'_y}{2} \Bigg]^2 F_j \cdot \left( \frac{2\pi T}{D}\right)^{2(\sqrt{K}-j\lambda/2)^2 - 2} 
  + \frac{1}{\hbar}  \Bigg[  \frac{J_E}{2} \Bigg]^2 \mu . \tag{S.12}
\end{equation}
with  $\mu = (1+\lambda^2/2) \sin(\pi \lambda^2/4) / (\pi \hbar v \sqrt{K})^2$. Comparing  Eqs.~(S.10) and (S.12) we once again see that $\gamma_0 \sim (J_x+J_y^{\prime})^2 T^{2(\sqrt{K}-\lambda/2)^2-1}$ and $\gamma_0^{\prime} \sim (J_x-J_y^{\prime})^2 T^{2(\sqrt{K}+\lambda/2)^2-1}$, and now we can also conclude that  $\tilde{\gamma}_0^E \sim J_E^2 T$ .

Thus we have obtained all rates appearing in the the conductance $\delta G(\omega)$ in Eq.~(S.4).


\begin{thebibliography}{10}

\bibitem{Konig} 
M. K\"onig {\em et al.}, Science {\bf 318}, 766 (2007).

\bibitem{Review}
For a review, see X.-L. Qi and S.-C. Zhang, Rev. Mod. Phys. {\bf 83}, 1057 (2011). 

\bibitem{Wu} C. Wu {\em et al.}, Phys. Rev. Lett. {\bf 96}, 106401 (2006).

\bibitem{Xu} C. Xu and J. E. Moore, Phys. Rev. B {\bf 73}, 045322 (2006).

\bibitem{Meidan}
D. Meidan and Y. Oreg, Phys. Rev. B 72, 121312(R) (2005).

\bibitem{Schmidt}
T.L. Schmidt {\em et al.}, Phys. Rev. Lett. {\bf 108} 156402 (2012).

\bibitem{Lezmy}
N. Lezmy {\em et al.}, Phys. Rev. B {\bf 85} 235304 (2012).

\bibitem{M}
J. Maciejko \textit{et al.}, Phys. Rev. Lett. {\bf 102} 256803 (2009).

\bibitem{T}
Y. Tanaka \textit{et al.}, Phys. Rev. Lett. {\bf 106} 236402 (2011).

\bibitem{WinklerBook}
R. Winkler, {\em Spin-Orbit Interaction Effects in Two-Dimensional Electron and Hole Systems}  
(Springer, Berlin, 2003).

\bibitem{Buhmann}
H. Buhmann, J. Appl. Phys. {\bf 109}, 102409 (2011).

\bibitem{BHZ}
B. A. Bernevig {\em et al.}, Science {\bf 314}, 1757 (2006).

\bibitem{Kane}
C. L. Kane and E. J. Mele, Phys. Rev. Lett. {\bf 95}, 226801 (2005).


\bibitem{Strom}
A. Str\"om {\em et al.}, Phys. Rev. Lett. {\bf 104}, 256804 (2010).

\bibitem{Meir}
Y. Meir and N. S. Wingreen, Phys. Rev. B {\bf 50}, 4947 (1994). 

\bibitem{Ujsaghy}
O. \'{U}js\'{a}ghy and A. Zawadowski, Phys. Rev. B {\bf 57}, 11598 (1998).

\bibitem{Zitko}
R. \v{Z}itko {\em et al.}, Phys. Rev. B {\bf 78}, 224404 (2008). 

\bibitem{VO}
J. I. V\"ayrynen and T. Ojanen, Phys. Rev. Lett. {\bf 106}, 076803 (2011).

\bibitem{Budich}
J. C. Budich {\em et al.}, Phys. Rev. Lett. {\bf 108}, 086602 (2012).

\bibitem{G}
T. Giamarchi, {\it Quantum Physics in One Dimension} (Oxford University Press, Oxford, 2003).

\bibitem{KaneFisher}
C. L. Kane and M. P. A. Fisher, Phys. Rev. B {\bf 46}, 15233 (1992).

\bibitem{M2}
J. Maciejko, Phys. Rev. B {\bf 85}, 245108 (2012).

\bibitem{Schuricht}
M. Pletyukhov and D. Schuricht, Phys. Rev. B {\bf 84}, 041309(R) (2011).

\bibitem{Feng}
X.-Y. Feng and F.-C. Zhang, J. Phys. Condens. Matter {\bf 23}, 105602 (2011).
  
\bibitem{Zitko2}
R. \v{Z}itko and J. Bon\v{c}a, Phys. Rev. B {\bf 84}, 193411 (2011).
 
\bibitem{Zarea}
M. Zarea {\em et al.}, Phys. Rev. Lett. {\bf 108}, 046601 (2012).
 
\bibitem{Isaev}
L. Isaev {\em et al.}, Phys. Rev. B {\bf 85}, 081107 (2012).

\bibitem{domefootnote}
The position of the ''dome'' in Fig. (1) is determined by the condition that the magnitude $|\tilde{J}_{E1}|$ of the scale-dependent
amplitude of the non-collinear term $\sim \sigma^{ y'}S^{z'}$ outgrows  $|\tilde{J}_x|$, $|\tilde{J}'_y|$, and $|\tilde{J}'_z|$  under RG.
Note that when $\theta > \pi/4$, this does not happen since now $J_y'=J_y\cos^2\theta+J_z\sin^2\theta$ will be large for large $J_z$,
making $|\tilde{J}'_y|$ dominate the RG flow. 

\bibitem{Malecki}
J. Malecki, J. Stat. Phys. {\bf 129}, 741 (2007).

\bibitem{Mahan}
G. Mahan, {\it Many-Particle Physics} (Kluwer Academic / Plenum Publishers, New York, 2000).

\bibitem{W}
X.-G. Wen, Phys. Rev. B {\bf 44} 5708 (1991).

\bibitem{Datta}
S. Datta {\em et al.}, 
Superlattice Microst. {\bf 1}, 327 (1985).

\bibitem{Furdyna}
J. K. Furdyna, J. Appl. Phys. {\bf 64}, R29 (1988).

\bibitem{Hinz}
J. Hinz {\em et al.}, Semicond. Sci. Technol. {\bf 21}, 501 (2006).

\bibitem{Hou}
C.-Y. Hou {\em et al.}, Phys. Rev. Lett. {\bf 102}, 076602 (2009).

\bibitem{Strom2}
A. Str\"om and H. Johannesson, Phys. Rev. Lett. {\bf 102}, 096806 (2009).

\bibitem{TeoKane}
J. C. Y. Teo and C. L. Kane, Phys. Rev B {\bf 79}, 235321 (2009). 

\bibitem{Efros} A. L. Efros and B. I. Shklovskii, {\em Electronic Properties of
Doped Semiconductors} (Springer, Heidelberg, 1989).

\bibitem{footnote}
Since $H_K^{\prime}$ in Eq. (\ref{HK}) is RG-relevant (marginally relevant) for $K<1$ $(K=1)$, the
condition $\delta I \ll G_0V$ for perturbation theory to be valid restrains $K$ to values
close to unity in the chosen voltage interval.

\bibitem{footnote2}
Note that a test requires the concurrent tunability of a back gate, so as to keep the electron density fixed \cite{Grundler}.

\bibitem{Grundler}
D. Grundler, Phys. Rev. Lett. {\bf 84}, 6074 (2000).

\bibitem{erratum}
The corrections in the present version, including the new Fig. 2, will be published in an Erratum to
Phys. Rev. B {\bf 86}, 161103(R) (2012).


\end{thebibliography}
\end{document}